\documentclass[12pt]{article}
\pdfoutput=1
\usepackage[utf8]{inputenc}
\usepackage[T1]{fontenc}
\usepackage{lmodern}
\usepackage{algorithm}
\usepackage{algpseudocode}
\usepackage{float}
\usepackage{amsmath}
\usepackage{amssymb}
\usepackage{mathtools}
\usepackage{graphicx}
\usepackage{array}
\usepackage{caption}
\usepackage[leftcaption,raggedright]{sidecap}
\usepackage[hidelinks]{hyperref}
\usepackage{slashed}
\usepackage{units}
\usepackage{booktabs}
\usepackage{url}
\usepackage{xspace}
\newcommand{\nuwro}{\textsc{NuWro}\xspace}
\newcommand\pubnumber{NuPhys2026-Rwik-Dharmapal-Banerjee}
\newcommand\pubdate{\today}
\def\affiliationA{Institute of Theoretical Physics, University of Wroc{\l}aw, Plac Maxa Borna 9,
50-204 Wroc{\l}aw, Poland}
\def\support{\footnote{This work was supported by the National Science Center under Grant
No.UMO-2021/41/B/ST2/02778. I am grateful to Artur M. Ankowski for
his guidance and supervision, and I also thank Jan T. Sobczyk, Krzysztof M. Graczyk, Jose L. Bonilla and H. Prasad for many fruitful discussions throughout this work.}}
\def\Title#1{\begin{center}{\Large #1}\end{center}}
\def\Author#1{\begin{center}{\sc #1}\end{center}}
\def\Address#1{\begin{center}{\it #1}\end{center}}

\newcommand\pubblock{
\rightline{
\begin{tabular}{l}
\pubnumber\\
\pubdate
\end{tabular}}
}
\newenvironment{Abstract}
{\begin{quotation}}
{\end{quotation}}
\newenvironment{Presented}
{\begin{quotation}
\begin{center}
PRESENTED AT
\end{center}
\bigskip
\begin{center}
\begin{large}}
{\end{large}
\end{center}
\end{quotation}}

\def\beq{\begin{equation}}
\def\eeq#1{\label{#1}\end{equation}}
\def\eeqn{\end{equation}}

\def\beqa{\begin{eqnarray}}
\def\eeqa#1{\label{#1}\end{eqnarray}}
\def\eeqan{\end{eqnarray}}

\let\bar=\overbar

\def\Dslash{\not{\hbox{\kern-4pt $D$}}}
\def\dslash{\not{\hbox{\kern-2pt $\del$}}}

\def\msb{{\bar{\ssstyle M \kern -1pt S}}}

\begin{document}

\begin{titlepage}
\pubblock
\vfill
\Title{A Consistent Treatment of Final-State Interactions in NuWro 
 Quasielastic Channel}
\vfill
\Author{RWIK DHARMAPAL BANERJEE\support}
\Address{\affiliationA}
\vfill

\begin{Abstract}
In this proceeding, I present a modified treatment of final-state interactions (FSI) in quasielastic (QE) lepton–nucleus scattering within the spectral-function (SF) framework of the \nuwro Monte Carlo generator. Our approach establishes a consistent correspondence between inclusive cross-section calculations and exclusive descriptions of hadron-propagation by combining a convolution-based formalism at the cross-section level with an event-level implementation in which interactions are classified as transparent or non-transparent within the \nuwro intranuclear cascade. This unified framework enables realization of FSI effects across inclusive observables and exclusive final states. We demonstrate the impact of this implementation by comparing predictions to both inclusive electron-scattering data and exclusive MicroBooNE measurements of CCQE-dominated observable, showing that the inclusion of FSI leads to a significant improvement in agreement with the data.
\end{Abstract}
\vfill

\begin{Presented}
NuPhys2026, Prospects in Neutrino Physics\\
King's College London, UK,\\
January 7--9, 2026
\end{Presented}

\vfill
\end{titlepage}

\def\thefootnote{\fnsymbol{footnote}}
\setcounter{footnote}{0}

\section{Introduction}
Accurate modeling of neutrino–nucleus interactions in the few-GeV energy region is essential for current and future neutrino-oscillation experiments \cite{Hyper-Kamiokande, dune}. In this regime, charged-current quasielastic (CCQE) scattering constitutes a dominant reaction channel, and its theoretical description is subject to significant uncertainties arising from nuclear effects. Among these, FSI—describing the propagation of the struck nucleon through the nuclear medium—play a crucial role in shaping both inclusive cross sections and exclusive final-state observables.

In this work, I focus on establishing a consistent treatment of FSI within the QE channel of the \nuwro generator \cite{sf_fsi}. The main idea is to build a direct mapping between the inclusive description of FSI—formulated in terms of a folding function—and its realization through the intranuclear cascade. This is achieved by tagging events as transparent or non-transparent, according to the nuclear transparency, and applying the corresponding modifications to the behavior of the struck nucleon's propagation.

This construction ensures that the underlying physics governing both inclusive observables and exclusive final states is consistent. The framework is tested against electron-scattering data and neutrino measurements from the MicroBooNE experiment, demonstrating its impact on improving the agreement with experimental results.

\section{\nuwro Monte Carlo Event Generator}
\nuwro is a Monte Carlo event generator \cite{SOBCZYK2005266} of lepton--nucleus interactions in the few-GeV energy region, developed at the University of Wroc{\l}aw since 2005. In addition to QE scattering, it includes hyperon production, single-pion production, deep-inelastic scattering, meson-exchange currents, electron--neutrino scattering, and coherent pion production. In the QE channel, \nuwro provides several options for modeling nuclear ground state such as different options of Fermi gas models or a SF approach. The current \nuwro version, tagged \texttt{25.11}, is publicly available on GitHub \cite{nuwro_code}.

\section{Spectral Function Approach}

In the SF approach, the nuclear ground state is described by the probability distribution of removing a nucleon with momentum $\mathbf{p}$ from the nucleus, leaving the residual with excitation energy E. This framework encodes both the shell structure of the nucleus and short-range correlations between nucleons, providing a more realistic description of bound nucleons in QE scattering compared to simplistic Fermi gas models.  

In \nuwro \texttt{25.11}, the SF approach is available for carbon, oxygen, argon, and iron. For carbon, oxygen, and iron, the implemented SFs are constructed by combining coincidence electron-scattering information on shell structure with theoretical calculations for nuclear matter in the local-density approximation. For argon, \nuwro employs \cite{PhysRevD.109.073004} the SFs extracted from the Jefferson Laboratory E12-14-012 measurements, which provided direct information on proton knockout and allowed for a state-of-the-art determination of the argon ground-state properties \cite{PhysRevC.99.054608, PhysRevD.105.112002, PhysRevD.107.012005}.    

\section{Final-State Interactions}
Within the SF framework, the inclusive QE cross section obtained in the plane-wave impulse approximation (PWIA) is modified by FSI through a convolution scheme with a folding function \cite{PhysRevC.44.2328}. The produced hadrons are then pushed to \nuwro's intranuclear cascade model \cite{PhysRevC.86.015505} for the description of possible rescatterings inside the residual nuclear medium.

\subsection{Convolution scheme}
The inclusive cross section with FSI is written as
\begin{equation}
\frac{d\sigma^{\mathrm{FSI}}}{d\omega d\Omega} = 
\int d\omega'\,
f_{\mathbf q}(\omega-\omega') \,
\frac{d\sigma^{\mathrm{PWIA}}}{d\omega' d\Omega}
\end{equation}
\vspace{0.5em}
where $\omega$ is the energy transfer, $\Omega$ denotes the solid angle of the outgoing lepton, and $f_{\mathbf q}$ describes the redistribution of strength induced by the interaction of the struck nucleon with the spectator system. 

The folding function is expressed as
\begin{equation}
f_{\mathbf q}(\omega) = 
\delta(\omega)\sqrt{T_A} + \left(1-\sqrt{T_A}\right)F_{\mathbf q}(\omega),
\end{equation}
where $T_A$ is the nuclear transparency \cite{Benhar:2003ka,PhysRevC.72.054602} and $F_{\mathbf q}(\omega)$ is a finite-width function responsible for the broadening of the cross section \cite{RevModPhys.80.189, PhysRevD.91.033005}. The first term in this equation corresponds to events in which the struck nucleon leaves the nucleus without initiating an intranuclear cascade, while the second term accounts for events in which rescattering must occur inside the nuclear medium.  

In addition to the broadening of the response, FSI induces a shift in the energy spectrum of the outgoing nucleon. This effect is incorporated through the real part of the optical potential $U_V$  \cite{PhysRevD.91.033005, PhysRevC.47.297}, as
\begin{equation}
f_{\mathbf q}(\omega-\omega')
\rightarrow
f_{\mathbf q}(\omega-\omega'-U_V)
\end{equation}  

\begin{figure}[t]
\centering
\begin{minipage}{0.48\textwidth}
\centering
\includegraphics[width=\textwidth]{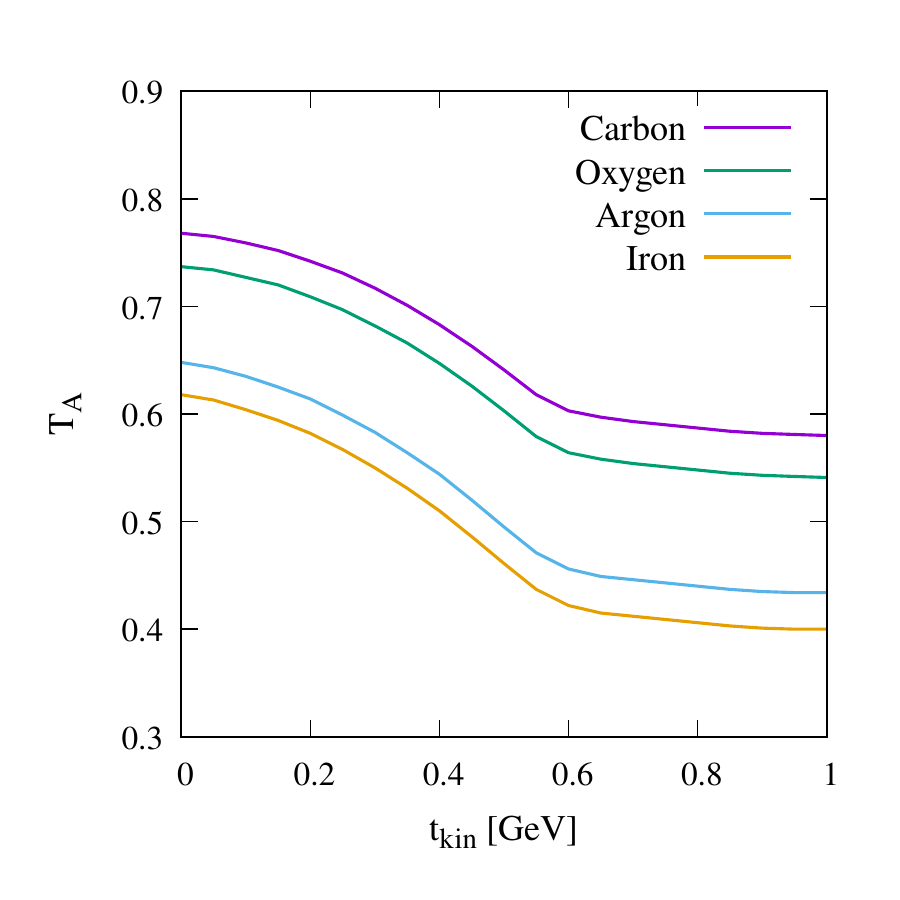}
\end{minipage}
\hfill
\begin{minipage}{0.48\textwidth}
\centering
\includegraphics[width=\textwidth]{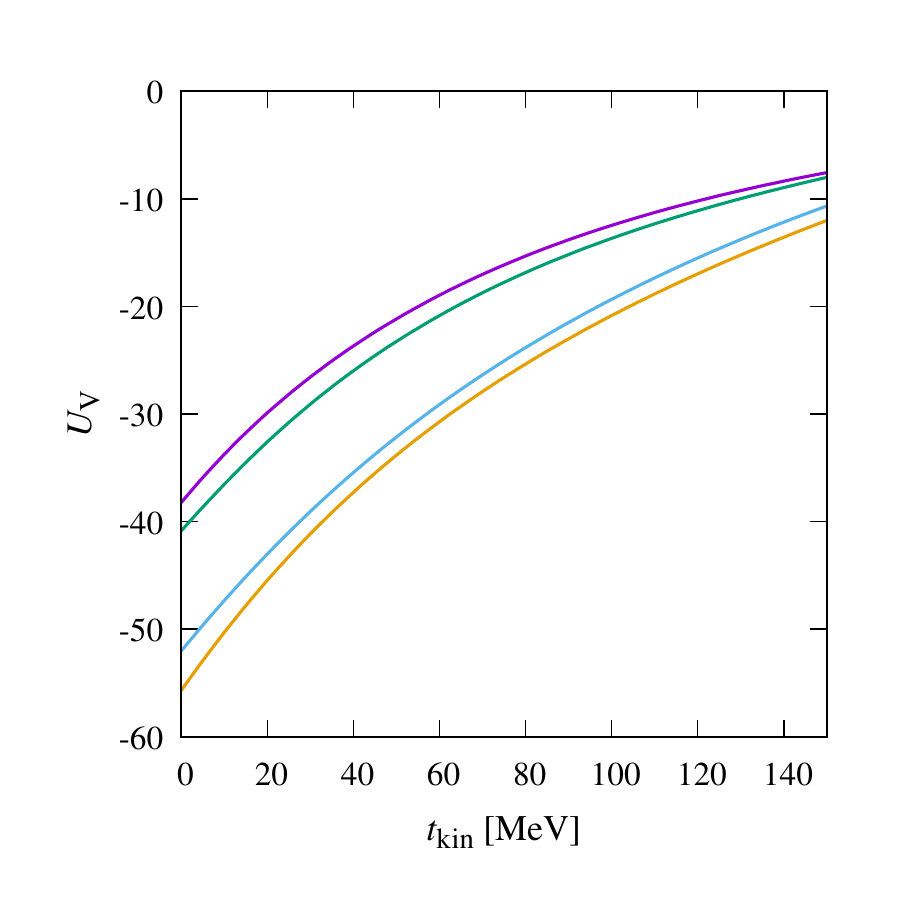}
\end{minipage}
\caption{Nuclear transparencies (left) and real parts of the optical potentials (right) implemented in \nuwro presented as a function of nucleon’s kinetic energy}
\label{fig:fsi_inputs}
\end{figure}

\subsection{\nuwro Intranuclear Cascade}
In \nuwro, the propagation of hadrons through the nuclear medium is described within a semi-classical intranuclear cascade model. After the primary lepton--nucleon vertex interaction, the produced hadron(s) are treated as quasi-free particle(s) moving through the nucleus and undergoing successive independent scatterings on bound nucleons. This description is well-justified if the transferred energies are large compared to typical nuclear binding energies and if the de Broglie wavelength of the propagating hadron is smaller than its mean free path in the medium. Within this picture, interference effects between different rescattering processes can be neglected, and the hadron trajectory is followed through a sequence of possible interaction points. 

The probability of reinteraction is determined locally from the nuclear density and the microscopic hadron--nucleon cross section. If a hadron traverses a short distance $\Delta x$, the probability that it survives without interaction is
\begin{equation}
\tilde p(\Delta x)=\exp\left(-\frac{\Delta x}{\lambda}\right),
\end{equation}
where the mean free path is given by $\lambda = 1/\rho \sigma$,
where $\rho$ is the nuclear density and $\sigma$ is the effective hadron--nucleon cross section. The distance $\Delta x$ traveled by a hadron between two successive possible interaction points is a fixed step size of 0.2 fm, over which the density is treated as approximately constant. At the end of each step, the code determines whether an interaction occurs according to $\tilde p$. These rescatterings modify the hadron kinematics and redistribute strength among different exclusive final states. 

\subsection{Algorithmic Implementation}
For QE events generated in the SF framework, the transparent
and non-transparent components of Eq.~2 are enforced at the event
level within the \nuwro cascade \cite{inprep}. The algorithm is briefly summarized below, where \(N_1\) is the struck nucleon and  \(N_2\) is its correlated partner nucleon for events tagged as correlated \cite{PhysRevC.53.1689}. In the following, by "standard cascade" I refer to the default \nuwro intranuclear cascade in which hadrons are propagated through the nuclear medium with interaction probabilities determined by the local density and hadron--nucleon cross sections, without imposing any additional criteria on the interaction history.

\noindent\rule{\textwidth}{0.4pt}
\begin{algorithmic}
\If{event is tagged transparent}
    \State propagate \(N_1\) without any rescattering 
    \If{event is tagged correlated}
        \State propagate \(N_2\) through the standard cascade 
    \EndIf
\EndIf

\If{event is tagged non-transparent}
        \State propagate \(N_1\) through the standard cascade once
        \State check whether \(N_1\) interacted
        \Repeat
        \If{\(N_1\) did not interact}
            \State discard the history associated with \(N_1\)
            \State redraw the propagation of \(N_1\)
            \State reduce the mean free path to increase interaction probability
            \State propagate \(N_1\) through cascade 
        \EndIf
    \Until{\(N_1\) interacts}

    \If{event is tagged correlated}
        \State propagate \(N_2\) through the standard cascade 
    \EndIf
\EndIf
\end{algorithmic}
\noindent\rule{\textwidth}{0.4pt}
\vspace{1em}
In this way, transparent events are comprehended as configurations in which the struck nucleon exits the nucleus undisturbed, whereas non-transparent events are those where the struck nucleon undergoes at least one intranuclear collision.

\begin{figure}[H]
\centering
\includegraphics[width=0.65\textwidth]{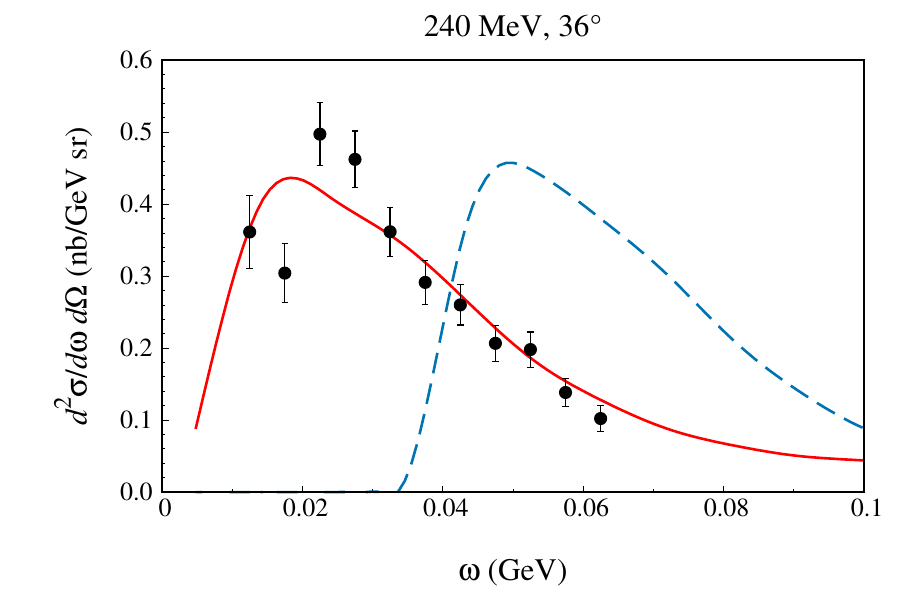}
\caption{Inclusive electron-carbon double-differential cross sections calculated within the SF approach in \nuwro. Results obtained with FSI (solid red line) are compared to calculations without FSI (dashed blue line) and to experimental
data. The experimental data is reported by Barreau et al. \cite{BARREAU1983515}. The panel is labeled according to beam energy and scattering angle.}
\label{fig:electron}
\end{figure}

\begin{figure}[H]
\centering
\includegraphics[width=0.7\textwidth]{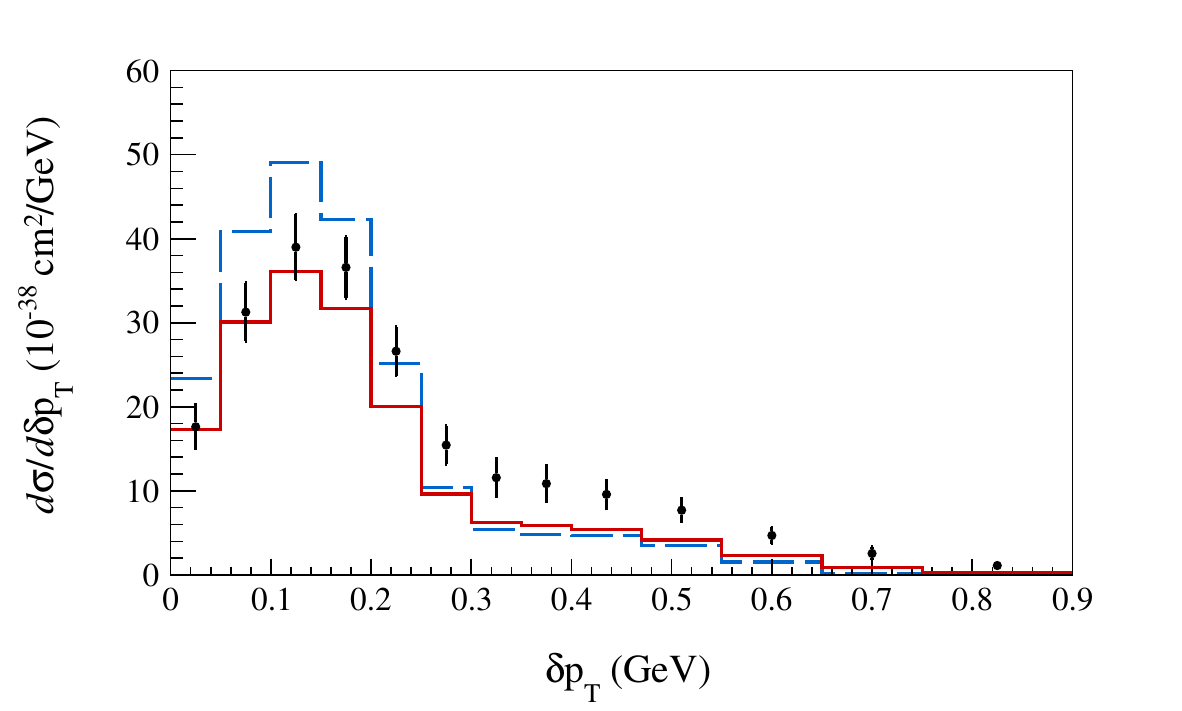}
\caption{MicroBooNE CC1p0$\pi$ double-differential cross sections  as a function of $\delta p_T$, compared with smeared \nuwro predictions with FSI (solid red line) and without FSI (dashed blue line).}
\label{fig:neutrino}
\end{figure}

\section{Results and Discussion}
I first present a validation of \nuwro prediction with inclusive electron scattering data in Fig.~\ref{fig:electron} The inclusion of FSI leads to a visible redistribution of strength, resulting in both broadening of the cross-section and shift of the QE peak. In contrast to the PWIA prediction, the full calculation agrees well with the data. At the exclusive level, MicroBooNE measurements of CC1p0$\pi$ interactions \cite{PhysRevLett.131.101802} are considered, focusing on an observable sensitive to nuclear effects. The transverse kinematic imbalance variable $\delta p_T$ is presented here. The comparison, shown in Fig.~\ref{fig:neutrino}, is performed after applying the additional smearing matrix provided by the experimental group. As seen, the inclusion of FSI leads to a pronounced reduction in the peak, resulting in a significantly improved agreement with the data. The deficit in the cross section at higher transverse momentum region suggests that the intranuclear cascade in \nuwro might need to be stronger.
\providecommand{\noopsort}[1]{}\providecommand{\singleletter}[1]{#1}%

\end{document}